\documentclass{aa}  

\usepackage{graphicx}
\usepackage{txfonts}
\usepackage{lipsum}
\usepackage{subcaption}         
\usepackage{lscape}             
\usepackage{placeins}           
                                
\usepackage{hyperref}

\begin{document}

   \title{Detection of periodic transit timing variations in warm sub-Saturn HD 332231 b}

   \subtitle{}


   \author{Gracjan Maciejewski\inst{1}\fnmsep\thanks{Corresponding author; \email{gmac@umk.pl}}
    }

   \institute{Institute of Astronomy, Faculty of Physics, Astronomy and Informatics,
              Nicolaus Copernicus University, Grudziadzka 5, 87-100 Toru\'n, Poland\\
    }

   \date{A\&A accepted, March 05, 2026}

  \abstract
   {Transit timing variations (TTVs) provide a powerful means to detect and characterise additional bodies in known planetary systems, even when they do not transit their host stars.}
   {We investigate the dynamical architecture of the HD~332231 system by analysing the TTVs of its close-in gas giant, HD~332231~b. Our goal is to assess whether the observed deviations from a linear ephemeris can be explained by the presence of an additional planetary companion.}
   {We refine the transit ephemeris of HD~332231~b using high-precision TESS photometry and complementary ground-based observations. We extract individual transit mid-times, construct an O--C diagram for transit timing data, and model the observed TTV signal through an extensive suite of N-body integrations covering a broad range of possible companion masses and orbital configurations.}
   {We detect a coherent TTV pattern with a period of approximately 6.7~years and an amplitude of about 45~minutes. Although numerous orbital configurations reproduce the observed TTVs, the combination of current radial velocity and photometric constraints yields a modest improvement in likelihood for solutions with an external planet on an orbit longer than 60~days, likely near a high-order mean-motion resonance and with moderate to high eccentricity.}
   {Our results suggest that HD~332231~b is part of a dynamically interacting multi-planet system. Continued transit monitoring and radial velocity follow-up will be essential to confirm the perturber's nature and refine the system's dynamical architecture.}


   \maketitle

\section{Introduction} \label{sect:introduction}

The HD~332231 system was first flagged as a promising candidate when the \textit{Transiting Exoplanet Survey Satellite} \citep[TESS,][]{2015JATIS...1a4003R} recorded a single transit with a depth of approximately 0.5\% ($\sim$5000~ppm) in stellar flux during its two-year Prime Mission \citep{2021ApJS..254...39G}. Subsequent high‑precision radial‑velocity (RV) observations obtained through the TESS‑Keck Survey \citep{2022AJ....163..297C} definitively confirmed the planetary nature of the signal \citep{2020AJ....159..241D}. These measurements revealed a companion of mass $80\,M_{\oplus}$ and radius $9.5\,R_{\oplus}$ on a nearly circular orbit with a period of 18.7~days. Designated HD~332231~b, the planet has a mean density of $0.5\, {\rm g\,cm^{-3}}$, placing it firmly in the category of warm sub-Saturnian worlds.

The host star is a 3 Gyr-old dwarf with a mass $M_{\star}$ of 1.2~$M_{\odot}$, radius $R_{\star}$ of 1.3~$R_{\odot}$, and effective temperature of 6130~K \citep{2023AJ....166...33M}. The study by \citet{2022AA...660A..99K} shows that the orbital angular momentum of HD~332231~b is aligned with the stellar spin. The orbital eccentricity was found to be consistent with zero within $2\sigma$ \citep{2020AJ....159..241D, 2022AA...660A..99K, 2024ApJS..272...32P} and must be primordial, because the tidal circularisation time scale is on the order of several hundreds of Gyr, much longer than the age of the system.

This orbital configuration of HD~332231~b supports quiescent formation scenarios, in which the planet was formed in-situ or ex-situ, beyond the water ice line, and smoothly migrated inward while embedded in a protoplanetary disk \citep[see][for a detailed review]{2018ARA&A..56..175D}. In such scenarios, any lower-mass planet, if formed together with HD~332231~b, might still be preserved. This finding motivated us to select the HD~332231 system for a project aimed at searching for additional planets accompanying gas giants in wide and circular orbits.

Furthermore, \citet{2022AA...660A..99K} reported possible variations in observed transit times, and \citet{2024ApJS..272...32P} flagged HD~332231~b as exhibiting significant transit timing variations (TTVs). Although these irregularities in orbital motion remained uncharacterised due to a limited number of observations, gravitational perturbations from an undetected planetary companion were invoked as a possible explanation \citep{2022AA...660A..99K}. These findings further boosted our motivation to revisit the system properties using TESS observations obtained over the last 5 years.

\section{TESS observations and data reduction} \label{sect:observations}

The apparent magnitude of HD~332231 in the TESS passband is $8\fm0417 \pm 0\fm0060$ \citep{2019AJ....158..138S}. This brightness ensures sub-millimagnitude precision on one-minute timescales in the photometric time series. The system was observed in sectors 14, 15, 41, 55, 75, 81, and 82 in short-cadence mode (120 second exposures). For sectors 75, 81, and 82, ultra-short-cadence data with 20-second resolution are also available. However, these were not incorporated into the final analysis, as their inclusion does not influence the results and a homogeneous data reduction procedure was preferred in all sectors. Details on individual observing runs are summarized in Table~\ref{table:tessRuns}.  

Fluxes were extracted from target pixel files (TPFs) using custom scripts based on the \texttt{Lightkurve} package \citep[version 2.5.0,][]{2018ascl.soft12013L}. Raw light curves, shown in Fig.~\ref{fig:tess_fluxes}, were produced via aperture photometry, employing the optimal masks provided within the TPFs. Scattered light was mitigated using the \texttt{RegressionCorrector} tool, which outperformed the standard sky-background subtraction based on median filtering. This approach reduced the out-of-transit (OOT) scatter by 2--3\% in Sectors 41--82 and by 13\% and 16\% in Sectors 14 and 15, respectively. To ensure that the final analysis relied solely on high-quality data, we discarded measurements with cadence quality flags other than zero.

The HD~332231~b transits were initially masked using the ephemeris of \citet{2022AA...660A..99K} and subsequently manually refined to account for variations in the transit timing. Long-term trends and low-frequency systematics were removed using the \texttt{flatten} function, which applies a Savitzky-Golay filter \citep{1964AnaCh..36.1627S} with a window width of 12 hours (361 data points). A $5\sigma$ clipping was then applied to the OOT data to eliminate outliers. Finally, the light curves were corrected for flux contamination from nearby stars using the \texttt{CROWDSAP} values provided by the pipeline for each sector.

\section{Modelling of TESS transit data} \label{sect:TESSmodeling}

Given the 18.7-day orbital period of HD~332231~b, typically only a single transit falls within each TESS sector. Exceptions include sectors 41 and 81, where two transits were recorded. Although the second event in sector 81 was only partially observed, our subsequent analysis confirmed its usefulness. In contrast, the transit in sector 14 was affected by scattered light, distorting its shape, and was excluded from the final analysis. Further discussion is provided in Appendix~\ref{app:sector14}.

We used the \texttt{Transit Analysis Package} \citep[\texttt{TAP},][]{2012AdAst2012E..30G} to derive the best-fitting transit model. TAP was applied to data segments centred on the mid-transit times. Each segment was five times the transit duration in length, ensuring a sufficient OOT baseline to accurately characterise photometric noise. This process was carried out in two iterations: the first used the initial ephemeris, and the second employed an updated ephemeris that includes a periodic term (see Sect.~\ref{sect:ephemeris}), enabling improved centring of the data chunks.

\texttt{TAP} generates analytic transit models following the formalism by \citet{2002ApJ...580L.171M}. The main model parameters describing transit geometry include the orbital inclination ($i_{\rm b}$), the scaled semi-major axis ($a_{\rm b}/R_{\star}$), and the planet-to-star radius ratio ($R_{\rm b}/R_{\star}$). Stellar limb darkening was modelled using a quadratic law \citep{1950HarCi.454....1K}, with linear ($u_1$) and non-linear ($u_2$) coefficients. These five parameters were jointly fitted using all available light curves. For each dataset, \texttt{TAP} also independently fitted the mid-transit time ($T_{\rm mid}$), uncorrelated and correlated photometric noise, and any remnant quadratic trends in the flux baseline. The orbital eccentricity ($e_{\rm b} = 0.029 \pm 0.019$) and argument of periastron ($\omega_{\rm b} = 39\degr \pm 31\degr$) were incorporated as Gaussian priors taken from \citet{2022AA...660A..99K} (the shadow model in their Table~2). 

We employed ten Markov Chain Monte Carlo (MCMC) chains, each with $10^6$ steps, to explore the posterior distributions of all free parameters. Because the posteriors of some parameters were significantly skewed, we adopted the mode (rather than the median) of the histograms (constructed following the Freedman-Diaconis rule) as the best-fit value. Uncertainties were estimated from the 68\% highest-density intervals surrounding the mode.

We also analysed posterior distributions for the transit impact parameter,
\begin{equation} 
b_{\rm b} = \frac{a_{\rm b}}{R_{\star}} \frac{1-e_{\rm b}^2}{1+e_{\rm b} \sin{\omega_{\rm b}}} \cos{i_{\rm b}} \, , 
\end{equation}
and the total transit duration,
\begin{equation} 
\tau_{\rm 14,b} = \frac{P_{\rm b}}{\pi} \frac{\sqrt{1-e_{\rm b}^2}}{1+e_{\rm b} \sin{\omega_{\rm b}}} \sqrt{\left( 1+\frac{R_{\rm b}}{R_{\star}} \right)^2 - b_{\rm b}^2} \, ,
\end{equation}
where $P_{\rm b}$ is the orbital period. Both quantities depend on $i_{\rm b}$ and $a_{\rm b}/R_{\star}$, which are correlated (see Fig.~\ref{fig:corners}). As discussed in Appendix~\ref{app:testFit}, a test run allowing transit parameters to vary independently across light curves revealed no statistically significant variations in these parameters.

\begin{figure}
    \includegraphics[width=\columnwidth]{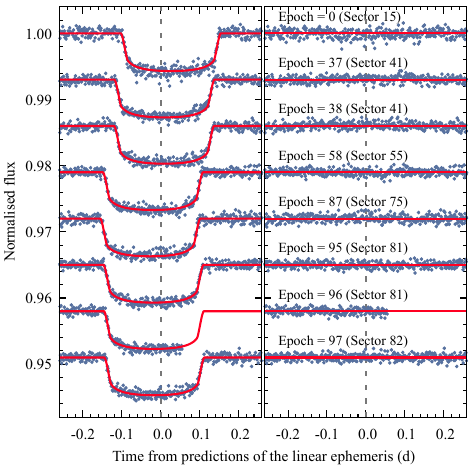}
    \caption{Transit light curves of HD~332231~b observed with
TESS in Sectors 15--82. Left: individual photometric time series sorted by the epoch of observation with numbering consistent with the refined ephemeris given in Sect.~\ref{sect:ephemeris}. Best-fitting models are overlaid in red. Right: corresponding photometric residuals from the transit models.}
    \label{fig:tessLCs}
\end{figure}

Individual light curves and their best-fitting models are shown in Fig.~\ref{fig:tessLCs}. The inferred transit parameters are reported in Table~\ref{table:results}.

\begin{table*}[h!]
\caption{\label{table:results}Transit parameters obtained for HD 332231~b}
\centering
\begin{tabular}{l l c c c}
\hline\hline
 & Parameter         & This paper & \citet{2020AJ....159..241D} & \citet{2022AA...660A..99K}\tablefootmark{a} \\
\hline
\multicolumn{5}{l}{Transit model}\\
& Orbital inclination, $i_{\rm{b}}$ $(\degr)$ & $89.97^{+0.02}_{-0.34}$ & $89.68^{+0.22}_{-0.28}$ & $89.78^{+0.22}_{-0.10}$\\
& Scaled semi-major axis, $a_{\rm b}/R_{\star}$ & $24.34^{+0.20}_{-0.37}$ & $24.21^{+0.62}_{-0.78}$ & $24.4^{+0.4}_{-0.3}$\\
& Radius ratio, $R_{\rm{b}}/R_{\star}$ & $0.07088^{+0.00039}_{-0.00031}$ & $0.06976^{+0.00041}_{-0.00039}$ & $0.0689 \pm 0.0003$\\
& Linear LD coefficient, $u_{\rm 1,TESS}$ & $0.246^{+0.069}_{-0.059}$ & $0.253 \pm 0.027$ & $0.261 \pm 0.014$ \\
& Quadratic LD coefficient, $u_{\rm 2,TESS}$ & $0.30 \pm 0.12$ & $0.289 \pm 0.034$ & $0.297 \pm 0.014$ \\
\multicolumn{5}{l}{Mid-transit times, $T_{\rm mid,b}$ $({\rm BJD_{TDB}}-2400000)$}\\
& ${\rm Epoch}=0$ (Sector 15) & $58729.68107 \pm 0.00049$ & $...$ & $...$ \\
& ${\rm Epoch}=37$ (Sector 41) & $59422.02742 \pm 0.00042$ & $...$ & $...$ \\
& ${\rm Epoch}=38$ (Sector 41) & $59440.73893 \pm 0.00042$ & $...$ & $...$ \\
& ${\rm Epoch}=58$ (Sector 55) & $59814.96045 \pm 0.00035$ & $...$ & $...$ \\
& ${\rm Epoch}=87$ (Sector 75) & $60357.61614 \pm 0.00044$ & $...$ & $...$ \\
& ${\rm Epoch}=95$ (Sector 81) & $60507.32448 \pm 0.00035$ & $...$ & $...$ \\
& ${\rm Epoch}=96$ (Sector 81) & $60526.03792 \pm 0.00073$ & $...$ & $...$ \\
& ${\rm Epoch}=97$ (Sector 82) & $60544.75165 \pm 0.00038$ & $...$ & $...$ \\
\multicolumn{5}{l}{Derived}\\
& Transit impact parameter, $b_{\rm{b}}$ $(R_{\star})$ & $0.012^{+0.141}_{-0.008}$ & $0.133^{+0.120}_{-0.092}$ & $0.10^{+0.04}_{-0.10}$\\
& Total transit duration, $\tau_{\rm{14,b}}$ (min) & $368.9^{+1.3}_{-1.1}$ & $369.4^{+1.6}_{-1.4}$ & $...$\\
\multicolumn{5}{l}{Transit ephemeris}\\
& Reference epoch, $T_{\rm 0,b}$\tablefootmark{b}  & $58729.6534^{+0.0054}_{-0.0066}$ & $58729.67987 \pm 0.00038$ & $58729.6814 \pm 0.0004$ \\
& Orbital period, $P_{\rm b}$ $({\rm d})$ & $18.71254^{+0.00015}_{-0.00012}$ & $18.71204 \pm 0.00043$ & $18.71205 \pm 0.00001$ \\
& TTV amplitude, $A_{\rm TTV}$ $({\rm d})$ & $0.0321^{+0.0052}_{-0.0040}$ & $...$ & $...$ \\
& TTV period, $P_{\rm TTV}$ $(E)$ & $131.0^{+8.9}_{-7.7}$ & $...$ & $...$ \\
& TTV phase, $\varphi_{\rm TTV}$ & $0.834^{+0.020}_{-0.021}$ & $...$ & $...$ \\
\hline
\end{tabular}
\tablefoot{
\tablefoottext{a}{Values taken from the planetary shadow model.}
\tablefoottext{b}{Given in ${\rm BJD_{TDB}}-2400000$.}
}
\end{table*}

\section{Refining the transit ephemeris for HD~332231~b} \label{sect:ephemeris}

The TESS observations yielded mid-transit times for eight epochs, spanning 97 orbital cycles -- equivalent to 1814 days, or nearly five years. A trial fit using a constant-period ephemeris in the classic form
\begin{equation}
     T_{\rm mid,b }(E) = T_{\rm 0,b} + P_{\rm b} \cdot E \, , \;
\end{equation}
where $E$ is the transit (cycle) number counted from the reference epoch $T_{\rm 0,b}$ (set to the mid-point of the transit observed in Sector 15), resulted in a reduced chi-squared ($\chi^2_{\rm red}$) value of approximately 440, revealing the residuals of the order of tens of minutes (see panel a of Fig.~\ref{fig:ttaov}). A periodogram analysis using the analysis of variance (AoV) algorithm \citep{1996ApJ...460L.107S} revealed a signal with a period of approximately 100 orbital cycles (panel b of Fig.~\ref{fig:ttaov}).

\citet{2022AA...660A..99K} previously used four mid-transit times from Sectors 14, 15, and 41 to refine the linear ephemeris. They reported that their spectroscopic transit observation, taken between Sectors 15 and 41, aligned with the model only when the mid-point was allowed to shift by around $+$20 minutes relative to the linear prediction. However, their transit timing dataset was too sparse to permit further interpretation. We incorporate this mid-transit time into our dataset, adopting the value derived from the planetary shadow model \citep[see Table 2 of][]{2022AA...660A..99K}. As described in Appendix~\ref{app:RMeffect}, we also attempted to determine a mid-transit time from the RV time series provided by \citet{2022AA...659A..44S}. Our re-analysis yielded an inconclusive result, and we therefore excluded it from further consideration.

To account for the timing variations, we expanded the ephemeris model by including a periodic term:
\begin{equation} \label{eq.eper}
     T_{\rm mid,b}(E) = T_{\rm 0,b } + P_{\rm b} \cdot E + A_{\rm TTV} \cdot \sin{\left[ 2 \pi \left( \frac{E}{P_{\rm TTV}} - \varphi_{\rm TTV} \right) \right]} \, , \;
\end{equation}
where $A_{\rm TTV}$, $P_{\rm TTV}$, and $\varphi_{\rm TTV}$ denote the amplitude, period, and phase of the TTV signal, respectively. These parameters were extracted from posterior distributions built using 100 Markov chains, each with $10^4$ steps (excluding the first 1000 as burn-in). The median values and uncertainties (15.9, 50, and 84.1 percentiles) are presented in Table~\ref{table:results}. The best-fitting model yields a $\chi^2_{\rm red}$ of 0.39, suggesting either slight over-fitting or that our timing uncertainties may be overestimated. In either case, the residuals leave no room for a more complex model at this stage.

\begin{figure*}
    \sidecaption
    \includegraphics[width=12cm]{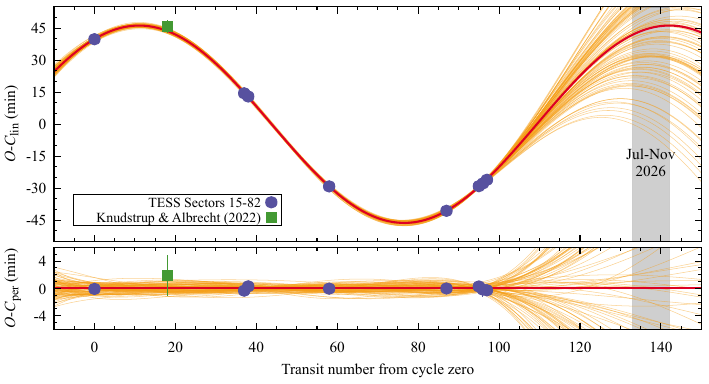}
    \caption{Transit timing variations (TTVs) of HD~332231~b. Top: the residuals after subtracting the linear ephemeris. The sinusoidal TTV model and 100 posterior models are shown. The grey shaded region marks the predicted 2026 observing window available from the ground. Bottom: final residuals after subtracting the periodic TTV model, provided by Eq.~\ref{eq.eper}.}
    \label{fig:ttoc}
\end{figure*}

Figure~\ref{fig:ttoc} presents the measured TTV signal, with the linear term of the ephemeris removed in upper panel, and the residuals after fitting the full model in lower panel. The best-fitting periodic model is plotted, along with 100 randomly drawn solutions from the posterior distribution to illustrate model's uncertainty. HD~332231~b exhibits TTVs with an amplitude of $46.3^{+7.4}_{-5.7}$ minutes -- over an order of magnitude larger than the uncertainties in the individual $T_{\rm mid,b}$ measurements -- and with a period of $131.0^{+8.9}_{-7.6}$ orbital cycles ($=2450^{+170}_{-140}$~days  $=6.7^{+0.5}_{-0.4}$~years).

Further follow-up observations are expected to improve the precision of the TTV parameters, particularly the period $P_{\rm TTV}$. The grey band in panels of Fig.~\ref{fig:ttoc} marks a transit window in 2026, observable from the ground. As the system is not scheduled for additional TESS coverage before August 2027\footnote{TESS-point Web Tool, \url{https://heasarc.gsfc.nasa.gov/wsgi-scripts/TESS/TESS-point_Web_Tool/TESS-point_Web_Tool/wtv_v2.0.py}, accessed on 2025 October 15}, ground-based transit monitoring could be valuable. While scheduling such observations is challenging due to the long $P_{\rm b}$ and $\tau_{\rm 14,b}$ of approximately 6 hours, even partial transits -- centred near ingress or egress -- can help constrain the ephemeris tighter.

\section{Search for additional transiting planets} \label{sect:search4planets}

The well-aligned, nearly edge-on, and circular orbit of HD~332231~b offers favourable conditions for detecting additional, smaller transiting planets -- provided they exist and share a similar coplanar alignment. To explore this possibility, we analysed the joint OOT photometric time series using the Analysis of Variance for planetary Transits method \citep[AoVtr,][]{2006MNRAS.365..165S}. This algorithm is a specialized adaptation of the classical AoV periodogram, tailored to identify box-shaped periodic signals characteristic of planetary transits.

To construct the OOT dataset, we removed the transits of HD~332231~b from the original light curves, applying a 15-minute margin around the predicted transit windows (based on the ephemeris in Eq.~\ref{eq.eper}). The light curves from all individual sectors were then combined, yielding a final dataset of 114,061 measurements.

We searched for periodic signals over trial periods ranging from 0.4 to 100 days, corresponding to orbital separations from roughly $2 \, R_{\star}$ outward. The frequency resolution was set to $2 \times 10^{-5}$ ${\rm d^{-1}}$. The AoVtr algorithm folds and bins the light curve for each trial period, fitting a negative-pulse model to identify potential transit-like events. Its sensitivity depends on both the number of phase bins and the transit duration. To optimize performance across the period range, we varied the bin count from 10 to 200 (in steps of 5) for periods shorter than that of HD~332231~b and from 80 to 400 (in steps of 10) for longer periods.

In each periodogram, we identified the strongest peak and evaluated its significance using the Signal Detection Efficiency (SDE) metric \citep{2000ApJ...542..257A,2002A&A...391..369K}, in the generalized form of \citet{2014A&A...561A.138O}. This metric quantifies the signal-to-noise ratio (S/N) in the power spectrum as: 
\begin{equation}
     {\rm S/R} = \frac{{\rm PWR} - {\rm PWR_{MED}}}{\sigma_{\rm MAD}} \, , \;
\end{equation}
where ${\rm PWR}$ is the peak power, ${\rm PWR_{MED}}$ is the median power in a frequency window around the peak, and $\sigma_{\rm MAD}$ is the noise level estimated via the median absolute deviation. As noted by \citet{2023ApJ...959L..16G}, there is no strict prescription for defining the width of the median filter. We set its width to one-fiftieth of the frequency span of the periodogram, providing a balance between smoothing the noise and preserving long-term trends. Injection-recovery tests (see below) indicated that signals with ${\rm S/N} > 30$ are likely to pass robust visual confirmation. 

The analysis returned no statistically significant signals indicative of additional transiting planets, either interior or exterior to HD~332231~b. To estimate the detection sensitivity of the current photometric dataset, we performed an injection-recovery analysis. Artificial box-shaped transits were inserted into the combined OOT light curve over periods from 0.4 to 100 days, incremented logarithmically (step size of 0.01 dex). We began with transit depths of 10 ppm, increasing by 10 ppm until the S/N exceeded the detection threshold. The injected transit duration was fixed at 2 hours. For each trial, the number of bins used by the AoVtr algorithm was calculated as the ratio of the trial period to the duration of the artificial transits, constrained between 10 and 400 bins. A periodogram was computed for each synthetic light curve, and the transit depth required to exceed an S/N of 30 was recorded and converted into a planetary radius limit using $R_{\star}$.

For interior orbits, our analysis was sensitive to planets as small as 0.8~$R_{\oplus}$ on the shortest-period orbits, increasing to 1.9~$R_{\oplus}$ for 10-day orbits. For exterior configurations, planets up to 2.5--3.0~$R_{\oplus}$ could be detected out to 35 days. Beyond this, data gaps between TESS sectors reduced detection efficiency, with limiting radii varying between 3 and 13 $R_{\oplus}$ depending on the extent of transit overlap with observed cadences.

This approach assumes that hypothetical planets maintain stable orbital periods, allowing for coherent phase-folding. However, TTVs comparable to or larger than the transit duration can smear the phase-folded signal and hinder detection. As discussed in Sect.~\ref{sect:interpretTTV}, the TTVs observed in HD~332231~b could be caused by an additional planetary companion. For near-circular orbits, the TTV amplitudes of two interacting planets are inversely proportional to their mass ratio and exhibit anti-correlated phases \citep{2016ApJ...821...96D,2016ApJ...823...72N}. If such a companion is smaller and less massive than HD~332231~b, its transits could exhibit TTVs with amplitudes significantly exceeding $A_{\rm TTV}$, complicating detection with our method.

To mitigate this, we reanalysed four shorter data subsets where the orbital period could be considered approximately constant. These subsets combined data from sectors 14+15, 41+55, 81+82, and 75+81+82. This alternative approach also yielded null results.

\section{Interpretation of the TTV signal} \label{sect:interpretTTV}

TTVs offer a powerful tool for detecting additional planets that gravitationally perturb the orbits of known transiting planets -- particularly when those companions are otherwise undetectable through other observational methods \citep{2005Sci...307.1288H,2005MNRAS.359..567A}. While early attempts to detect TTVs focused on massive, short-period planets using ground-based observations, it was the advent of space-based photometry -- most notably from the Kepler mission -- that enabled the first practical applications of this technique \citep{2010Sci...330...51H,2011ApJ...743..200B}. However, interpreting TTVs remains a complex inverse problem: different orbital architectures can produce similar timing signals, making it difficult to uniquely constrain the properties of the perturbing body.

Moreover, not all TTVs are necessarily planetary in origin. Other astrophysical mechanisms may also cause variations in transit timing. In the following subsections, we explore both scenarios: non-planetary explanations are considered in Sect.~\ref{sect:interpretTTVa}, while possible planetary perturbations are discussed in Sect.~\ref{sect:interpretTTVb}.

\subsection{Non-planetary scenarios} \label{sect:interpretTTVa}

The observed variations in the orbital period of HD~332231~b could be apparent rather than physical, potentially arising from non-planetary phenomena such as the light travel time effect (LTTE). In the presence of a third body, the barycentre of the star--planet pair would oscillate around the system's common centre of mass. This geometric effect would manifest as periodic shifts in transit timing, with the orbital period of the third body matching the observed $P_{\rm TTV}$. To produce a signal with an amplitude of $A_{\rm TTV}$, the radial displacement of the system would need to be approximately $A_{\rm TTV} \cdot c \approx 5.6 \, {\rm au}$, corresponding to a RV amplitude of around $25 \, {\rm km \, s^{-1}}$.

The longest RV time series is reported by \citet{2020AJ....159..241D} and was acquired with the Levy Spectrograph on the 2.4 m Automated Planet Finder telescope \citep[APF,][]{2014SPIE.9145E..2BR,2014PASP..126..359V} at Lick Observatory. It covers only about 6\% of the TTV signal's phase. On one hand, these observations may have missed such significant RV variations, particularly near the extrema of the signal. This observational limitation is consistent with the absence of long-term RV trends noted by \citet{2020AJ....159..241D}. On the other hand, to generate the required LTTE signal, the mass of the hypothetical companion would need to be at least $5.6 \, M_{\odot}$ -- significantly more massive than HD~332231 itself. This makes the scenario implausible because spectroscopic analyses show no evidence of a secondary stellar component \citep{2020AJ....159..241D,2023AJ....166...33M}.

We also examined the possibility of a gravitationally bound brown dwarf or stellar companion in a wide orbit, constituting a hierarchical triple system and capable of producing the observed $P_{\rm TTV}$. Using the formalism of \citet{2011A&A...528A..53B} and assuming circular orbits for both HD~332231~b and the hypothetical companion, we found no viable solutions that could reproduce the observed $A_{\rm TTV}$. Even under the most favourable configuration -- mutually perpendicular orbital planes and a companion mass of $50 \, M_{\rm Jup}$ -- the predicted TTV amplitude was only half the observed value.

Another potential explanation involves apsidal precession of an eccentric orbit of HD~332231~b. In this scenario, the variations in $P_{\rm b}$ could be an apparent effect arising from the precession of the planetary orbit. The required eccentricity, estimated as $2 \pi \, A_{\rm TTV} / P_{\rm b} \approx 0.01$, lies within the $2\sigma$ confidence interval reported by \citet{2024ApJS..272...32P}. We calculated precession timescales considering contributions from general relativity, as well as tidal and rotational deformations of the host star and planet, using the equations of \citet{2009ApJ...698.1778R}. For these calculations, we adopted a stellar tidal Love number of $k_{2,\star} \approx 0.0087$ \citep{1995A&AS..109..441C}, and a planetary value of $k_{\rm 2,b} = 0.39$ reported for Saturn \citep{2017Icar..281..286L}. Assuming a planetary rotation period of 10 hours -- consistent with Solar System gas giants -- we found that the shortest precession periods, driven by relativistic and rotational effects, are on the order of $10^5$ years. Tidal effects lead to even longer timescales, exceeding $10^7$ years. These are vastly inconsistent with the observed TTV period, ruling out precession as a viable explanation.

Lastly, we considered magnetic activity cycles as a possible cause of timing variations. In such cases, cyclic changes in the host star's gravitational potential -- induced by shape deformations linked to magnetic activity -- can modulate the orbital period of a companion \citep{1992ApJ...385..621A}. Using Eq.~(13) from \citet{2010MNRAS.405.2037W}, together with the stellar parameters from \citet{2023AJ....166...33M} and a rotation period inferred from the projected equatorial velocity \citep{2022AA...660A..99K}, we found the expected timing variations to be just $\sim10^{-2}$ seconds. This is five orders of magnitude smaller than the observed $A_{\rm TTV}$. Even when adopting less conservative assumptions regarding the stellar quadrupole moment \citep{1998MNRAS.296..893L}, the effect remains negligible.

Given the implausibility of these non-planetary explanations -- whether geometric, dynamical, or magnetic in nature -- we are led to consider alternative causes. Planetary perturbations thus emerge as the most likely origin of the observed TTV signal.

\subsection{Planetary perturbations} \label{sect:interpretTTVb}

To explore whether gravitational perturbations from an undetected planetary companion could explain the observed TTVs, we performed a suite of reconnaissance simulations using the \texttt{TTVFast} code \citep{2014ApJ...787..132D}. It employs a numerically efficient symplectic integrator to compute transit times in simulated planetary systems. Synthetic mid-transit times were used to construct O--C (observed minus calculated) diagrams, with transit ephemerides refined via linear regression. We then analysed the residuals for sinusoidal patterns using the AoV algorithm, searching over periods from 2 to 7000 epochs at a frequency resolution of $1 \times 10^{-4}$ epoch$^{-1}$.

We considered perturbers with an orbital period $P_{\rm c}$ ranging from 3 to 350 days (in steps of $2 \times 10^{-4}$ days), eccentricity $e_{\rm c}$ from 0 to 0.85 (step size 0.05), argument of periastron from 0$^\circ$ to 330$^\circ$ (in steps of 30$^\circ$), and a mass $M_{\rm c}$ of 5, 10, 25, and 50 $M_{\oplus}$. The mass of HD~332231~b was calculated by adopting the RV amplitude $K_{\rm b} = 17.5 \pm 1.1$ ${\rm m\, s^{-1}}$ from \citet{2022AA...660A..99K} and the stellar mass $M_{\star} = 1.16 \pm 0.03 \, M_{\odot}$ from \citet{2023AJ....166...33M}. 

Each simulation spanned 10,000 days -- approximately four full TTV cycles -- with integration steps of 0.025 days for inner configurations and 0.1 days for outer ones. Due to the system's alignment, only prograde, coplanar orbits were considered. Configurations resulting in orbit crossings were excluded. For each model, we assessed whether the simulated TTV period matched $P_{\rm TTV}$ within uncertainties and whether the required perturber mass to reproduce $A_{\rm TTV}$ remained below the RV detection threshold defined as: 
\begin{equation}
     K_{\rm lim} = \sqrt{2} \, \sigma_{\rm jitter,HIRES} \approx 4.3 \, {\rm m \, s^{-1}} \, , \;
\end{equation}
where $\sigma_{\rm jitter,HIRES}$ represents the RV jitter in the high-precision HIRES \citep{1994SPIE.2198..362V} dataset acquired from Keck I by \citet{2020AJ....159..241D}. For the purposes of this exercise, we approximated the RV signature of the perturber as a sinusoid for simplicity.

\begin{figure*}
    \sidecaption
    \includegraphics[width=12cm]{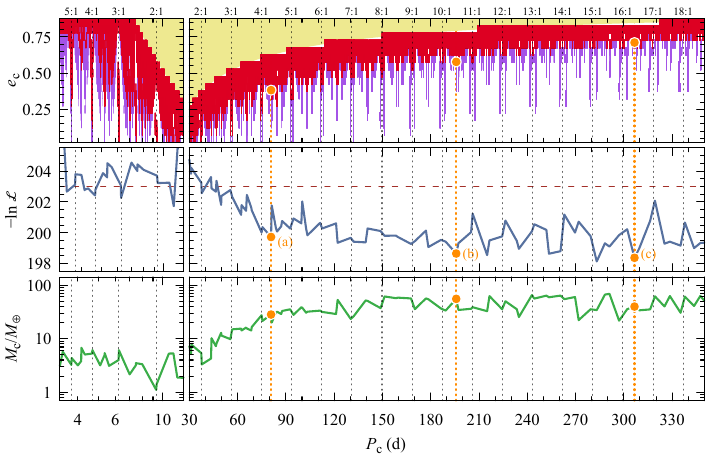}
    \caption{Exploration of potential planetary perturbers responsible for the observed TTV signal in the HD~332231 system. Upper panels: Map of simulated configurations in the parameter space of perturber's orbital period and eccentricity. Red pixels denote solutions that reproduce both the observed TTV period and amplitude while remaining below the RV detection threshold. Violet pixels indicate configurations that match the TTV signal and may be confirmed with the current RV data. Yellow regions correspond to dynamically unstable (orbit-crossing) configurations. Middle panels: Negative log-likelihood metric, $- \ln \mathcal{L}$, as a function of the perturber's orbital period for various resonance zones. The dashed horizontal line indicates the $- \ln \mathcal{L}$ value for a single-planet model. Bottom panels: Perturber mass required to reproduce the observed $A_{\rm TTV}$ across the tested orbital periods. Orange points and dotted vertical lines indicate the three illustrative solutions presented in Fig.~\ref{fig:models}.}
    \label{fig:map}
\end{figure*}

Many configurations successfully reproduced $P_{\rm TTV}$, typically clustering near low- and high-order orbital resonances. These are visualized in Fig.~\ref{fig:map} (upper panels), with red pixels indicating models that match both $P_{\rm TTV}$ and $A_{\rm TTV}$, but produce RV signals below $K_{\rm lim}$, hence cannot be confirmed with the Doppler data. Violet pixels represent configurations matching the TTV signal but requiring perturber masses that would provide detectable RV amplitudes. Yellow regions mark dynamically unstable configurations with crossing orbits.

To assess whether the existing RV data could constrain the perturber's properties more precisely, we attempted to identify the best-fitting two-planet solutions near inner and outer $x$:$1$, $x$:$2$, and $x$:$3$ resonances, where $x$ denotes relevant integer values across the explored period range. Here, the notation $p$:$q$ refers to the period ratio of inner ($p$) to outer ($q$) planets. We combined the \texttt{TTVFast} algorithm with the dynamic nested sampling routine from the \texttt{dynesty} package \citep{2020MNRAS.493.3132S} to explore the likelihood function in its general form:
\begin{equation}
 \label{Eq:logL}
  \ln \mathcal{L} = -\frac{1}{2} \sum_i \left[ \frac{(\mathcal{O}_i - \mathcal{M}_i)^2}{\sigma_i^2} + \ln(2\pi\sigma_i^2) \right]  \, ,
\end{equation}
where $\mathcal{O}_i$ and $\sigma_i$ are the observed values and uncertainties, $\mathcal{M}_i$ are the corresponding model predictions, and the sum runs over all data points. In addition to the transit timing dataset for HD~332231~b, the observational data were supplemented with the OOT RV data taken from \citet{2020AJ....159..241D}, comprising 13 high-precision HIRES observations and 68 moderate-precision APF measurements. Data from the SONG telescope, also reported by those authors, were excluded due to their large uncertainties, which were comparable to the RV amplitude induced by HD~332231~b. All timestamps were converted to ${\rm BJD_{TDB}}$ to ensure consistent timing for TTV analysis. 

The model included 14 free parameters: the orbital period, mass, eccentricity, argument of periastron, and mean anomaly for both HD~332231~b and the trial planet, as well as the offsets and jitter terms for each of the two RV datasets (added in quadrature to the formal uncertainties). Priors were assigned uniformly within limits optimised individually based on our reconnaissance simulations. For each considered resonance, we explored solutions slightly below and above the exact orbital period commensurability. We employed 7000 live points, allowing the parameter space to be sampled until the change in log-evidence fell below $\Delta \ln \mathcal{Z}$ of $0.01$. The solutions with the highest $\ln \mathcal{L}$ were subsequently fine-tuned using the Nelder--Mead minimisation routine implemented in \texttt{SciPy} \citep{Virtanen2020}, driven by the negative log-likelihood-based cost function:
\begin{equation}
 \label{Eq:chi2}
     \mathcal{C} = -2\ln \mathcal{L}  \, .
\end{equation}
The variation of $-\ln \mathcal{L}$ with the perturber's orbital period is shown in the middle panels of Fig.~\ref{fig:map}. The dashed line indicates the single-planet baseline, for which the TTV signal was approximated by a pure sinusoid, as in Eq.~\ref{eq.eper}.

In our approach, the RVs primarily constrain the planetary masses and orbital parameters, whereas the transit data tightly define the TTV signature. Relative to the single-planet solution, two-planet configurations with orbital periods $P_{\rm c} \lesssim 60$~days do not improve the fit. As shown in the lower panels of Fig.~\ref{fig:map}, this is primarily because the plausible perturbers are low-mass (1--10 $M_{\oplus}$, with a median of 5 $M_{\oplus}$), producing radial velocity amplitudes of only 0.2--3.3 ${\rm m \, s^{-1}}$ (median 1 ${\rm m \, s^{-1}}$). Near the 2:1 resonances, even Earth-mass perturbers are capable of reproducing the observed TTV signal, but their induced Doppler signals fall below current detection thresholds. Similarly, as demonstrated in Sect.~\ref{sect:search4planets}, the TESS photometry lacks the sensitivity to detect such perturbers unless they are low-density, gaseous planets with radii larger than 1.5 $R_{\oplus}$ (for inner orbits) or 3 $R_{\oplus}$ (for outer orbits). Rocky planets in these regimes remain undetectable with current instrumentation.

\begin{figure*}
    \sidecaption
    \includegraphics[width=12cm]{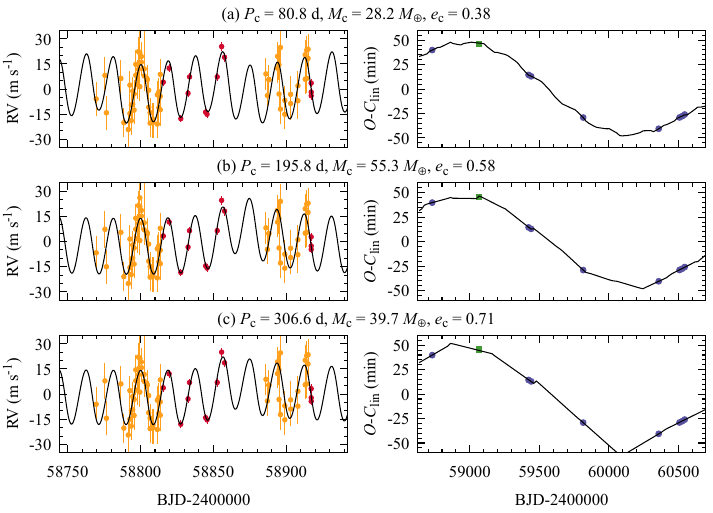}
    \caption{Comparison of three illustrative two-planet dynamical models fitted to HD~332231. Each row corresponds to a different perturber configuration with orbital periods of about 81 days (a), 196 days (b), and 307 days (c); the corresponding masses and orbital eccentricities are also provided. The left panels show the modelled RV signals overlaid on HIRES and APF data points (red and orange points, respectively). The right panels present the corresponding TTV model compared to observed mid-transit times. By analogy to Fig.~\ref{fig:ttoc}, blue points comes from TESS light curves, and the green square is from spectral observations by \citet{2022AA...660A..99K}.}
    \label{fig:models}
\end{figure*}

For longer orbital periods, that is $P_{\rm c} \gtrsim 60$ days, we observed a modest improvement in $\ln \mathcal{L}$, forming a shallow plateau for $P_{\rm c} \gtrsim 120$ days. Although no long-term trend is evident in the full RV dataset \citep{2020AJ....159..241D}, a linear trend of $0.19 \pm 0.04$ ${\rm m \, s^{-1} \, day^{-1}}$ appears in the HIRES data from BJD 2,458,815 to 2,458,858. Three additional HIRES measurements near BJD 2,458,917.15 deviate from this trend, suggesting possible curvature that could be indicative of an underlying periodic signal. In Fig.~\ref{fig:models}, we present three illustrative solutions for a perturber orbiting near the 13:3 resonance (a), 21:2 resonance (b), and 33:2 resonance (c), also indicated in Fig.~\ref{fig:map}. In each case, the model RV curves reproduce the local HIRES trend, while the TTV curves match the observed mid-transit times. The TTVs become increasingly non-sinusoidal with longer $P_{\rm c}$, featuring characteristic flat segments associated with enhanced perturbations near periastron. However, the statistical preference for these long-period configurations over more compact solutions remains limited.

Although our systematic transit search returned no detections, we additionally inspected both de-trended and raw TESS light curves, searching for monotransits, as expected for solutions such as those presented in Fig.~\ref{fig:models}. For example, model (a) predicts two transit windows (in Sectors 14 and 41), while models~(b) and~(c) each predict a single window (in Sectors 14 and 55, respectively). Although transits in these cases are effectively ruled out, long-period transits could still evade detection due to incomplete photometric coverage.

\section{Conclusions} \label{sect:conclusions}

Our analysis reveals that HD~332231~b exhibits strong periodic TTVs with a cycle of approximately 6.7 years and an amplitude of about 45 minutes. We provide an updated transit ephemeris that includes a periodic term, which will be important for planning future transit observations. Dynamical simulations suggest that the observed TTVs can be explained by gravitational perturbations from an additional, as-yet undetected planetary companion. Although many orbital configurations can reproduce the TTV signal, current photometric and RV data are insufficient to uniquely constrain its properties. However, the RV data show a modest improvement in likelihood for solutions with an outer companion on an orbit longer than about 60~days, though the statistical preference over more compact configurations remains limited.

The system is not expected to be revisited by TESS until August 2027, limiting access to space-based transit data in the near term. In the meantime, ground-based transit observations could significantly improve constraints on the TTV period. Likewise, acquiring new high-precision RV measurements would help test the plausibility of outer perturber scenarios, particularly those near high-order mean-motion resonances. If such a companion exists, our modelling indicates that it would likely occupy a relatively eccentric orbit -- strikingly different from the inner planet's dynamically calm configuration.

HD~332231 thus remains a compelling target for continued monitoring. Coordinated efforts involving ground-based transit observations, enhanced RV coverage, and refined dynamical modelling
could yield valuable insights into the formation and evolution of multi-planet systems featuring TTVs and eccentric outer perturbers.

\begin{acknowledgements}
I would like to thank the anonymous referee for constructive and insightful comments, which helped improve the robustness of the analysis. I acknowledges the financial support from the National Science Centre, Poland, through grant no. 2023/49/B/ST9/00285. This paper includes data collected with the TESS mission, obtained from the MAST data archive at the Space Telescope Science Institute (STScI). Funding for the TESS mission is provided by the NASA Explorer Program. STScI is operated by the Association of Universities for Research in Astronomy, Inc., under NASA contract NAS 5-26555. This research made use of \texttt{Lightkurve}, a Python package for Kepler and TESS data analysis \citep{2018ascl.soft12013L}. This research has made use of the SIMBAD database and the VizieR catalogue access tool, operated at CDS, Strasbourg, France, and NASA's Astrophysics Data System Bibliographic Services.
\end{acknowledgements}

\bibliographystyle{aa} 
\bibliography{aa57788-25} 

\begin{appendix}
\onecolumn
\section{TESS observations} \label{app:tabsfigs}

\begin{table*}[h!]
\caption{\label{table:tessRuns}Details on TESS observations used in this study}
\centering
\begin{tabular}{cccccccc}
\hline\hline
Cycle & Sector & Camera  & CCD & from -- to & $N_{\rm data}$ & PNR & $N_{\rm tr}$  \\
      &        &         &     &           &                & (ppm~min$^{-1}$) &   \\
\hline
 2 & 14 & 1 & 1 & 2019 Jul 18 -- 2019 Aug 14 & 18424 & 826 & 0 \\
 2 & 15 & 1 & 2 & 2019 Aug 15 -- 2019 Sep 10 & 17749 & 883 & 1 \\
 4 & 41 & 1 & 1 & 2021 Jul 23 -- 2021 Aug 20 & 18322 & 644 & 2 \\
 4 & 55 & 2 & 2 & 2022 Aug 05 -- 2022 Sep 01 & 18882 & 633 & 1 \\
 6 & 75 & 4 & 3 & 2024 Jan 30 -- 2024 Feb 26 & 19474 & 631 & 1 \\
 6 & 81 & 2 & 1 & 2024 Jul 15 -- 2024 Aug 10 & 16048 & 578 & 2 \\
 6 & 82 & 2 & 2 & 2024 Aug 10 -- 2024 Sep 05 & 18138 & 578 & 1 \\
\hline
\end{tabular}
\tablefoot{Cycle is a predefined period during which TESS observes different regions of the sky: the southern ecliptic hemisphere in odd-numbered cycles and the northern ecliptic hemisphere in even-numbered cycles. Camera and CCD specify a sensor used. $N_{\rm data}$ is the number of useful measurements in a final light curve. PNR is the photometric noise rate in parts per million (ppm) of the normalised flux per minute of exposure \citep[see][]{2011AJ....142...84F}, calculated for out-of-transit data points. $N_{\rm tr}$ is the number of transits qualified for this study.}
\end{table*}

\begin{figure*}[h!]
	\includegraphics[width=\columnwidth]{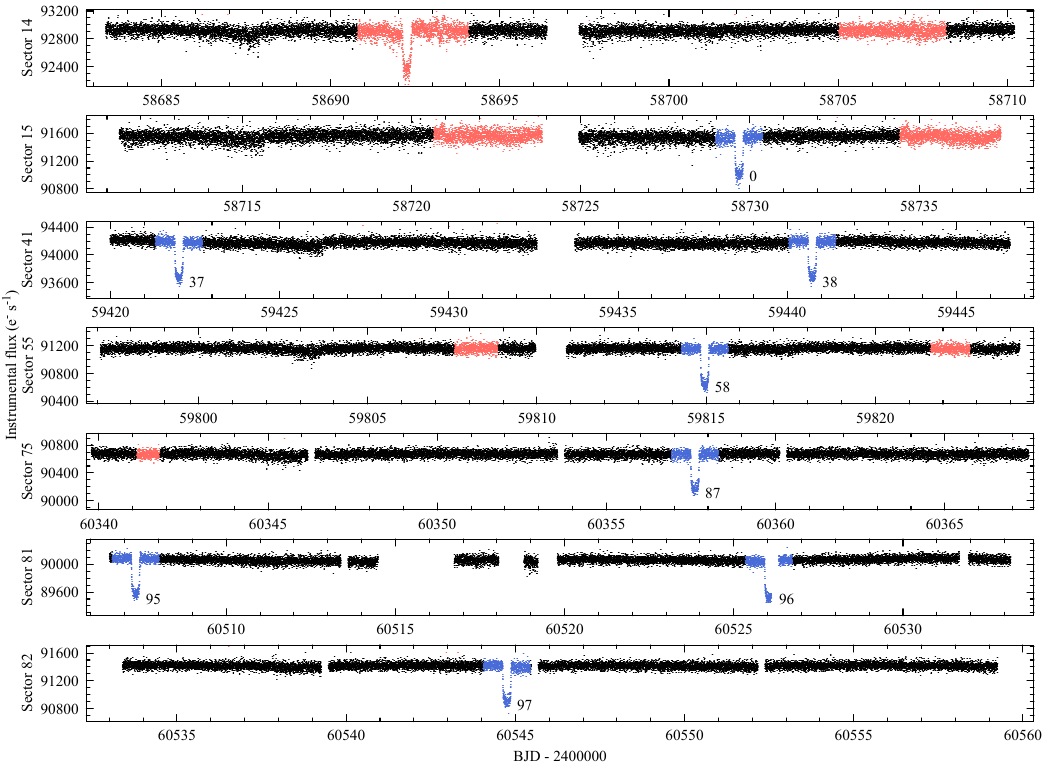}
    \caption{TESS instrumental light curves of HD 332231 in individual observing sectors. Black points represent photometric measurements with quality flag ${\rm QF} = 0$, which were included in the final analysis. Red points correspond to measurements with ${\rm QF} > 0$ that were excluded. The segments centred on transits of planet b used in transit modelling (Sect.~\ref{sect:TESSmodeling}), are shown in blue, with individual transits marked by their epoch numbers.}
    \label{fig:tess_fluxes}
\end{figure*}

\newpage

\section{Modelling of the transit light curve in Sector 14} \label{app:sector14}

\citet{2020AJ....159..241D} identified the transit of HD~332231~b in Sector~14 data that had been flagged as affected by scattered light and were therefore excluded from the original Pre-search Data Conditioning Simple Aperture Photometry (PDC-SAP) analysis. Post-transit observations in this sector also display signs of an Argabrightening event, with erratic flux variations near BTJD~1693.2 -- approximately seven hours after fourth contact. To recover these compromised data, \citet{2020AJ....159..241D} employed a standard aperture photometry approach using \texttt{Lightkurve}, augmented with an advanced systematic correction based on spacecraft quaternion time series within each exposure. The resulting Sector~14 light curve was then included in their analysis and contributed to the refinement of the transit ephemeris. 

\begin{figure}[h!]
    \includegraphics[width=\columnwidth]{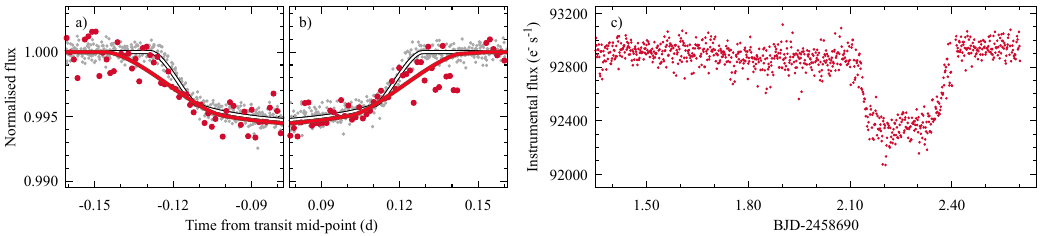}
    \caption{Transit of HD~332231~b in Sector~14. Panels~a and~b compare the Sector~14 transit light curve (red points) with the phase-folded data from other sectors (grey points), zoomed in on ingress and egress to highlight discrepancies. The corresponding best-fitting models are shown as red and black/white lines. Panel~c shows the raw instrumental flux around the Sector~14 transit, illustrating systematics affecting the measurements.}
    \label{fig:tessS14}
\end{figure}

However, visual inspection of Fig.~4 in \citet{2020AJ....159..241D} suggests that the Sector~14 transit appears anomalously long. To further investigate this anomaly, we reprocessed the Sector~14 data using our pipeline, deliberately bypassing the quality flag criteria and retaining all available exposures. A preliminary fit performed with \texttt{TAP} indicates that the Sector~14 transit is $66^{+11}_{-48}$ minutes longer and $1300^{+300}_{-900}$ ppm deeper compared to transits observed in other sectors. Although these differences are not statistically significant (at the $\sim$1.5$\sigma$ level), they raise concerns regarding the reliability of this dataset. Panels~a and~b of Fig.~\ref{fig:tessS14} illustrate the Sector~14 transit overlaid on the phase-folded light curves from other sectors, along with their respective best-fitting models, clearly showing the discrepancy. As shown in panel~c, the raw instrumental flux exhibits a distinct offset in the out-of-transit baseline, accompanied by mild undulations and clusters of negative outliers.

We conducted a suite of tests, including masking groups of outlying data points and applying relative flux offsets to correct for potential discontinuities between segments. These attempts did not yield satisfactory improvements, suggesting that the underlying systematic effects are more complex than assumed. This exercise ultimately demonstrated that the quality of the transit light curve in Sector~14 renders it unsuitable for detailed analysis. Nevertheless, the initial detection of the transit in this sector played a crucial role in constraining $P_{\rm b}$ and establishing the preliminary transit ephemeris by \citet{2020AJ....159..241D}.

\newpage

\section{TAP modelling posterior distributions} \label{app:TAPcorners}

\begin{figure*}[h!]
	\includegraphics[width=\columnwidth]{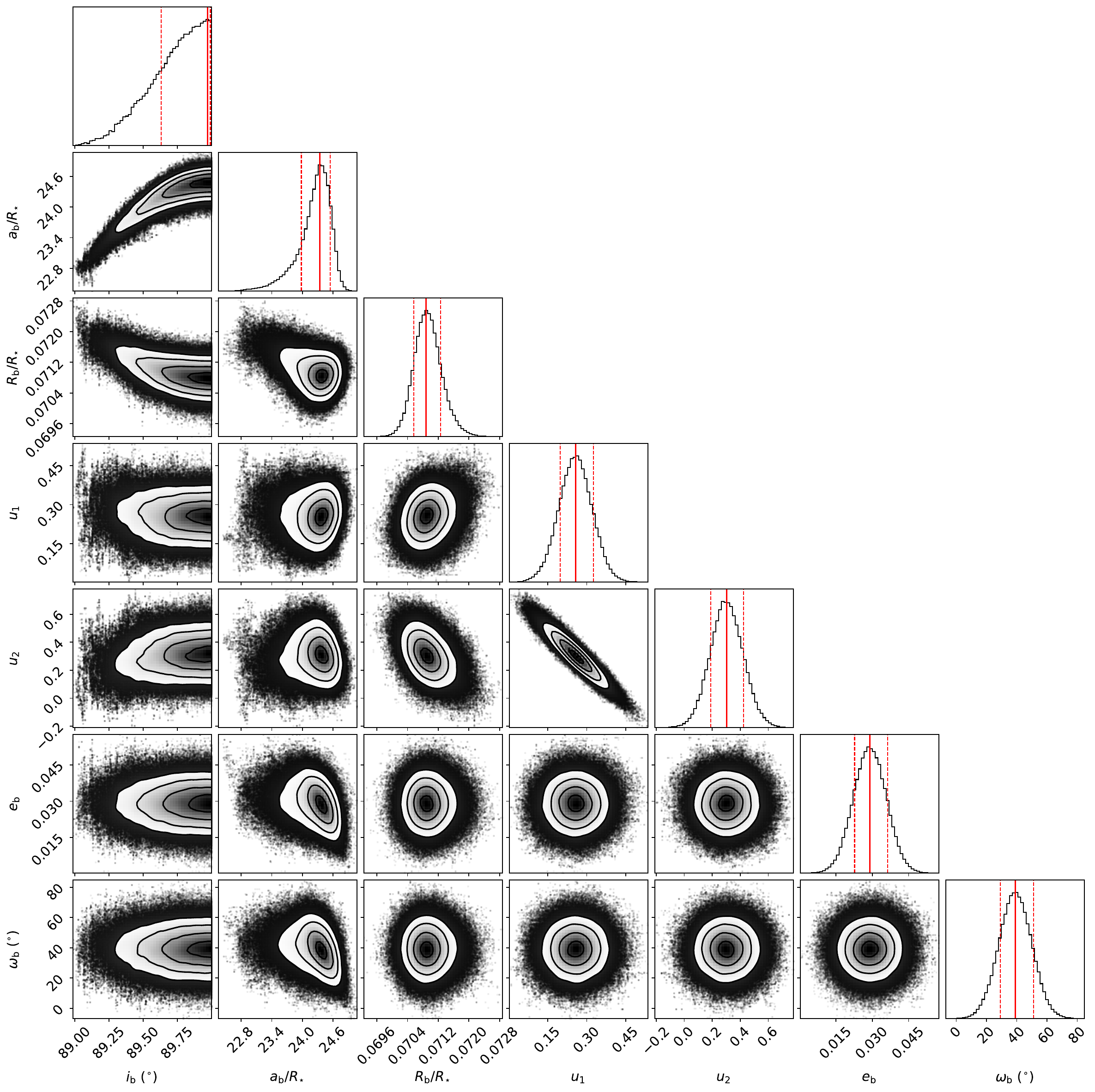}
    \caption{Corner plot showing the posterior distributions of the fitted parameters for the transit model of HD~332231~b, based on the TESS data. The diagonal panels show marginalized 1D distributions, while the off-diagonal panels display joint 2D projections to illustrate parameter covariances. Contours correspond to 1$\sigma$, 2$\sigma$, and 3$\sigma$ confidence levels. The orbital parameters $e_{\rm b}$ and $\omega_{\rm b}$ were sampled under Gaussian priors taken from \citet{2022AA...660A..99K}. The fit was performed using the TAP model described in Sect.~\ref{sect:TESSmodeling}, and informed the dynamical simulations discussed in later sections.}
    \label{fig:corners}
\end{figure*}

\newpage

\section{Search for evolution of the transit parameters} \label{app:testFit}

\begin{figure*}[h!]
    \sidecaption
    \includegraphics[width=12cm]{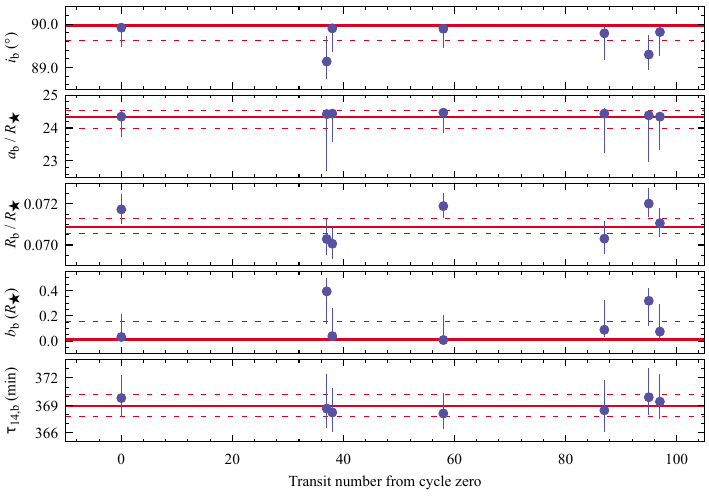}
    \caption{Parameters for individual complete transits observed with TESS. The best-fitting values for the joint model and their 1$\sigma$ uncertainties, reported in Sect.~\ref{sect:TESSmodeling}, are marked with horizontal continuous and dashed lines, respectively.}
    \label{fig:parvar}
\end{figure*}

In a test run, we allowed $i_{\rm b}$, $a_{\rm b}/R_{\star}$, and $R_{\rm b}/R_{\star}$ to be independently fitted to the individual transit light curves. We considered only the complete ones to base the subsequent analysis on the most reliable data that limited our sample to 7 light curves. The remaining free parameters were set as in the model described in Sect.~\ref{sect:TESSmodeling}. The results are collected in Table~\ref{table:indLCs} together with derived individual values for $b_{\rm b}$ and $\tau_{\rm 14,b}$. As illustrated in Fig.~\ref{fig:parvar}, these transit parameters do not reveal any variation correlated with time, represented by transit epochs $E$.  

\begin{table*}[h!]
\caption{\label{table:indLCs}Best-fitting parameters for individual complete transits of HD~332231~b}
\centering
\begin{tabular}{ccccccc}
\hline\hline
Sector & $E$ & $i_{\rm b}$ $(\degr)$ & $a_{\rm b}/R_{\star}$ & $R_{\rm b}/R_{\star}$ & $b_{\rm b}$ ($R_{\star}$) & $\tau_{\rm 14,b}$ (min)\\
\hline
15 &  0 & $89.921^{+0.054}_{-0.450}$ & $24.35^{+0.16}_{-0.62}$ & $0.07173^{+0.00075}_{-0.00070}$ & $0.032^{+0.183}_{-0.022}$ & $369.8^{+2.5}_{-2.1}$ \\
41 & 37 & $89.139^{+0.579}_{-0.402}$ & $24.42^{+0.14}_{-1.73}$ & $0.07029^{+0.00095}_{-0.00077}$ & $0.393^{+0.105}_{-0.255}$ & $368.6^{+3.8}_{-2.1}$ \\
41 & 38 & $89.905^{+0.065}_{-0.546}$ & $24.44^{+0.15}_{-0.86}$ & $0.07006^{+0.00083}_{-0.00073}$ & $0.039^{+0.221}_{-0.027}$ & $368.2^{+2.6}_{-2.2}$ \\
55 & 58 & $89.897^{+0.071}_{-0.444}$ & $24.46^{+0.13}_{-0.63}$ & $0.07189^{+0.00062}_{-0.00057}$ & $0.008^{+0.199}_{-0.006}$ & $368.1^{+2.2}_{-1.7}$ \\
75 & 87 & $89.791^{+0.143}_{-0.606}$ & $24.44^{+0.16}_{-1.19}$ & $0.07031^{+0.00088}_{-0.00075}$ & $0.089^{+0.236}_{-0.061}$ & $368.5^{+3.3}_{-2.4}$ \\
81 & 95 & $89.301^{+0.455}_{-0.360}$ & $24.38^{+0.14}_{-1.40}$ & $0.07201^{+0.00076}_{-0.00066}$ & $0.318^{+0.105}_{-0.201}$ & $369.9^{+3.2}_{-2.0}$ \\
82 & 97 & $89.820^{+0.123}_{-0.562}$ & $24.35^{+0.15}_{-1.02}$ & $0.07106^{+0.00073}_{-0.00065}$ & $0.075^{+0.222}_{-0.051}$ & $369.4^{+3.0}_{-1.9}$ \\
\hline
\end{tabular}
\end{table*}

\newpage

\section{Preliminary transit timing analysis} \label{app:OCperiodogram}

\begin{figure*}[h!]
	\includegraphics[width=\columnwidth]{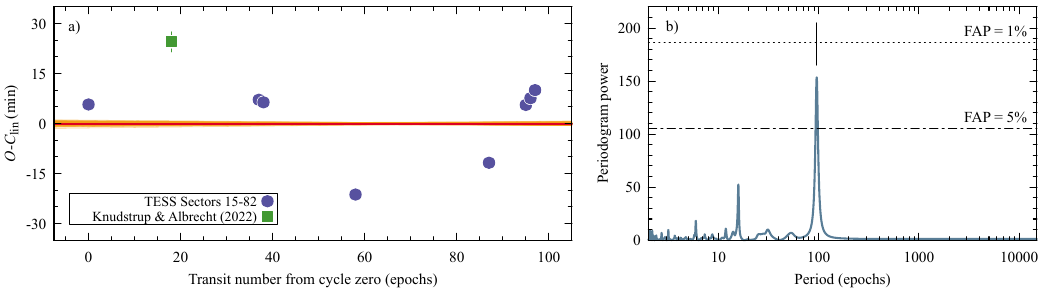}
    \caption{Results of the preliminary analysis of the transit-timing dataset. Panel~a: Transit-timing residuals relative to a trial linear ephemeris. The symbol coding follows that in Fig.~\ref{fig:ttoc} and is explained in the legend. Timing uncertainties for the TESS data are smaller than the symbol size. The uncertainty of the trial ephemeris is illustrated by 100 posterior realisations, shown in orange. Panel~b: AoV periodogram of the residuals obtained after subtracting the refined linear ephemeris shown in panel~a. Horizontal lines indicate the false-alarm probability (FAP) levels, estimated using a bootstrap procedure applied to 100,000 resampled datasets. The strongest peak, near 96 epochs, has a FAP of 1.9\% and is marked by a vertical line. Several harmonics of this signal are also visible.}
    \label{fig:ttaov}
\end{figure*}

\section{Revisiting the Rossiter-McLaughlin effect of HD~332231~b} \label{app:RMeffect}

In their study, \citet{2022AA...660A..99K} observed a transit of HD~332231~b on UT 2020 August 5 using the High Accuracy Radial velocity Planet Searcher for the Northern hemisphere \citep[HARPS-N,][]{2012SPIE.8446E..1VC} on the 3.6 m Telescopio Nazionale Galileo. The 6.4-hour observing run began near first contact and concluded shortly after last contact. The Doppler data showed a characteristic distortion due to the Rossiter-McLaughlin (RM) effect. Three modelling techniques were applied to extract the sky-projected obliquity $\lambda_{\rm b}$ -- the angle between the projected stellar spin axis and the planet's orbital axis. The most precise method, which analysed spectral line distortions, yielded $\lambda_{\rm b} = -2 \pm 6$ degrees, indicating that the planet's orbital plane and the stellar equator are aligned within $1\sigma$.

In contrast, \citet{2022AA...659A..44S} reported a significantly different result, based on a transit observed on UT 2020 October 18 using the CARMENES high-resolution spectrograph \citep{2014SPIE.9147E..1FQ} on the 3.5 m telescope at Calar Alto Observatory. Their observations spanned from shortly before second contact to just after fourth contact. They also collected RV measurements on the nights before and after the transit, each lasting 1.75 hours, to better constrain the RV baseline. Their analysis produced $\lambda_{\rm b} = -42 \pm 11$ degrees, suggesting a moderate misalignment.

\citet{2022AA...660A..99K} attributed this discrepancy to \citeauthor{2022AA...659A..44S}'s use of a linear ephemeris based solely on TESS photometry from Sectors 14 and 15, along with contemporaneous RV data. Indeed, for that epoch, $E = 22$, their ephemeris differed by 14 minutes from that given in Eq.~\ref{eq.eper} (Sect.~\ref{sect:ephemeris}). To improve on this, \citeauthor{2022AA...660A..99K} incorporated two additional mid-transit times from TESS sector 41 and introduced a free parameter, $\Delta T_{\rm 0,b}$, to account for a potential mid-transit time offset. Their best-fitting model yielded $\Delta T_{\rm 0,b}$ of $\sim$20 minutes, detected with $>5\sigma$ significance. When they fixed $T_{\rm 0,b}$ to their updated ephemeris, that is, adopting $\Delta T_{\rm 0,b} = 0$, the derived obliquity shifted to $\lambda_{\rm b} = -31 \pm 6$ degrees, consistent with \citeauthor{2022AA...659A..44S}'s value. This finding underscored the risk of obtaining spurious $\lambda_{\rm b}$ estimates when transit timing variations are neglected. \citet{2022AA...660A..99K} also questioned the higher projected rotation velocity, $v \sin i_{\star} = 16.3^{+6.9}_{-4.4}$ ${\rm km \; s^{-1}}$, reported by \citet{2022AA...659A..44S}, noting its strong correlation with $\lambda_{\rm b}$ \citep[see][]{2011ApJ...738...50A}. 

To revisit these results, we reanalysed both in-transit RV datasets using the \texttt{allesfitter} package \citep{allesfitter-code, allesfitter-paper}, employing the ephemeris with the periodic term from Eq.~\ref{eq.eper}. We additionally incorporated the OOT RV data reported by \citet{2020AJ....159..241D}. This dataset comprises 13 high-precision HIRES observations obtained with HIRES at Keck I \citep{1994SPIE.2198..362V}, together with 68 moderate-precision measurements acquired using APF Levy Spectrograph on the 2.4 m Automated Planet Finder telescope \citep{2014SPIE.9145E..2BR,2014PASP..126..359V}. We excluded the data from the SONG telescope, as their large uncertainties -- comparable to the planetary RV semi-amplitude -- rendered them unsuitable for our analysis.

In a trial run, we fitted the transits from Sectors 15--82 to independently determine transit parameters. The results agreed with those from TAP (Sect.~\ref{sect:TESSmodeling}) within $1\sigma$, and we subsequently adopted the TAP-derived values as Gaussian priors on $R_{\rm b}/R_{\star}$, the normalized sum of radii $(R_{\rm b}+R_{\star})/a_{\rm b}$, and the inclination through $\cos i_{\rm b}$. Free parameters in our model included $\lambda_{\rm b}$, the RV semi-amplitude $K_{\rm b}$, the orbital eccentricity $e_{\rm b}$ and argument of periastron $\omega_{\rm b}$ represented by $\sqrt{e_{\rm b}} \cos{\omega_{\rm b}}$ and $\sqrt{e_{\rm b}} \sin{\omega_{\rm b}}$, as well as instrumental offsets and jitter terms. Following \citet{2022AA...660A..99K}, we modelled stellar limb darkening using a quadratic law with coefficients $u_1$ and $u_2$, adopting nominal values interpolated from \citet{2011A&A...529A..75C} for the $V$ band. These were allowed to vary with Gaussian priors \citep[$\sigma = 0.1$ for $u_1$, and $0.2$ for $u_2$;][]{2022AJ....163..228P}. The alternative reparametrization, based on the $q$-space \citep{2013MNRAS.435.2152K}, yielded consistent results, so we retained the physical $u$-form to match our TAP modelling. The space of free parameters was explored with dynamic nested sampling with 500 live points and default settings \citep{allesfitter-paper}.

\begin{figure*}
    \sidecaption
    \includegraphics[width=12cm]{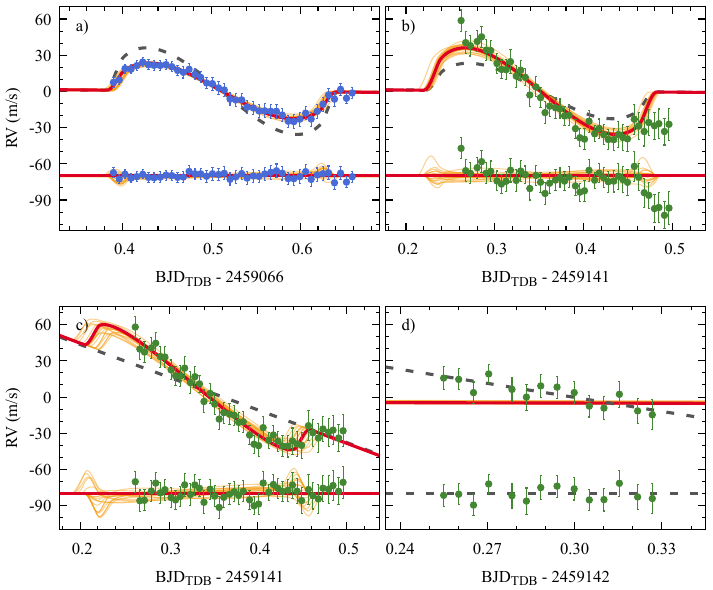}
    \caption{Rossiter-McLaughlin effect observed during two transits of HD~332231~b. Panel a: HARPS-N observations from \citet{2022AA...660A..99K}, showing the in-transit RV anomaly consistent with a well-aligned orbit. Blue points denote the measured RVs with 1$\sigma$ uncertainties. The solid red line shows the best-fitting model, while the orange-shaded curves represent 20 posterior samples illustrating the range of plausible solutions. The dashed gray line indicates the higher-amplitude RM model based on the CARMENES data. Panel b: CARMENES observations from \citet{2022AA...659A..44S}, shown as green points with the corresponding best-fitting model. The dashed gray line shows the HARPS-N-based solution for comparison. Panel c: The same CARMENES data, re-fitted with an added linear RV trend (dashed gray line), yielding a reduced RM amplitude. Panel d: Out-of-transit CARMENES measurements taken the night after the transit, showing a non-zero RV slope, also indicated with a dashed gray line. The RV changes induced by the orbital motion of HD~332231~b are shown in red and remain negligible in this scale.}
    \label{fig:rm}
\end{figure*}

We attempted to jointly model the RM signals from both HARPS-N and CARMENES. However, the resulting fits were unsatisfactory. Panels a and b of Fig.~\ref{fig:rm} show clearly that the RM amplitudes differ markedly. Since RM amplitude scales approximately as $(R_{\rm b}/R_{\star})^2 \sqrt{1-b_{\rm b}^2} \, (v \sin i_{\star})$ \citep{2018haex.bookE...2T}, and the planetary/stellar radii and impact parameter must be identical, this discrepancy implies a difference in $v \sin i_{\star}$. As such, we proceeded to model the transits separately.

Although $v \sin i_{\star}$ can, in principle, be derived from RM data alone, central transits (low $b_{\rm b}$ or $\cos i_{\rm b}$ near zero) introduce a degeneracy between $v \sin i_{\star}$ and $\lambda_{\rm b}$, degrading parameter precision \citep{2011ApJ...738...50A}. Our tests confirmed this effect for HD~332231~b, where the posterior distribution of $v \sin i_{\star}$ showed a long high-velocity tail inconsistent with spectral line measurements. We therefore adopted a Gaussian prior of $v \sin i_{\star} = 5.3 \pm 1.0$ ${\rm km \, s^{-1}}$, based on Keck-HIRES spectra \citep{2020AJ....159..241D}.

Key results are shown in Table~\ref{table:rmeffect}. All other parameters agreed with literature values within uncertainties and are omitted for brevity. The best-fitting models (solid lines) and 20 posterior draws (shaded lines) are shown in Fig.~\ref{fig:rm}. To highlight differences in the RM amplitudes, dashed lines show the alternative model in each panel.

In both transits, $\lambda_{\rm b}$ is consistent with $0^{\circ}$, supporting alignment and confirming findings of \citet{2022AA...660A..99K}. Our HARPS-N value of $v \sin i_{\star}$ agrees within 0.8--1.1 $\sigma$ with the values of $5.3 \pm 1.0$, $5.4 \pm 1.0$, and $7.0 \pm 0.5$ ${\rm km \, s^{-1}}$, derived from spectral analysis by \citet{2020AJ....159..241D}. It also agrees within 0.9--1.5 $\sigma$ with the values of $5.64 \pm 0.14$, $5.63 \pm 0.11$, and $5.89^{+0.12}_{-0.13}$ ${\rm km \, s^{-1}}$, obtained using three methods by \citet{2022AA...660A..99K}.

However, our CARMENES-derived $v \sin i_{\star} = 10.0 \pm 0.6$ ${\rm km \, s^{-1}}$ is higher by 4--7$\sigma$, suggesting potential systematics. The residuals in panel b of Fig.~\ref{fig:rm} reveal outliers at the end of the CARMENES series. \citet{2022AA...659A..44S} masked the final four out-of-transit points, citing increased airmass but without identifying a cause. We explored this further by fitting a model with a linear baseline trend ("CARMENES+slope", panel c of Fig.~\ref{fig:rm}). This model yielded $\lambda_{\rm b} = 0^{+21}_{-20}$ degrees and a lower $v \sin i_{\star} = 6.3 \pm 0.7$ ${\rm km \, s^{-1}}$, consistent with other datasets. The out-of-transit RVs taken one day later with CARMENES (panel d of Fig.~\ref{fig:rm}) also show a slope of $-27.4 \pm 9.2$ ${\rm m \, s^{-1} \, day^{-1}}$, differing from zero at nearly 3$\sigma$, which supports using a baseline slope in the RM model. 

Mid-transit times from our RM analysis are also listed in Table~\ref{table:rmeffect}. For HARPS-N, our result matches \citet{2022AA...660A..99K} within $1\sigma$ and is more precise. Nevertheless, we use their value in the transit timing analysis, as it was obtained via the planetary shadow method, which provides robust results.

For CARMENES, the situation is more complex. Although including a baseline trend reduced the RM amplitude, it doubled the timing uncertainty and shifted $T_{\rm mid,b}$ by 26 minutes ($\approx 2.2 \sigma$ effect). Given these uncertainties, we excluded the CARMENES data from our timing analysis. Actual in-night trends may have a more complex nature, and a refined reduction of the original observations may help clarify this issue.

\begin{table*}[h!]
\caption{\label{table:rmeffect}Best-fitting parameters from the RM effect modelling for HD~332231~b}
\centering
\begin{tabular}{lccc}
\hline\hline
Parameter         & HARPS-N & CARMENES & CARMENES+slope\\
\hline
$\lambda_{\rm b}$ (deg) & $-4^{+20}_{-19}$ & $1.2^{+8.2}_{-8.0}$ & $0^{+21}_{-20}$\\
$v \sin i_{\star}$ (${\rm km \; s^{-1}}$) & $6.2^{+0.5}_{-0.4}$ & $10.0 \pm 0.6$ & $6.3 \pm 0.7$\\
$T_{\rm mid,b}$ (${\rm BJD_{TDB}} - 2459000$) & $66.5114^{+0.0018}_{-0.0018}$ & $141.3498 ^{+0.0041}_{-0.0042}$ & $141.3316^{+0.0070}_{-0.0077}$\\
\hline
\end{tabular}
\end{table*}

\end{appendix}
\end{document}